\documentclass[submission,copyright,creativecommons]{eptcs}
\providecommand{\event}{WWV 2014} 
\usepackage{breakurl}             
\usepackage{graphicx}
\usepackage{amsthm}\theoremstyle{definition}
\usepackage{bussproofs}
\newtheorem{mydef}{Definition}
\newtheorem{proposition}{Proposition}
\usepackage{syntax}
\usepackage{listings}

\begin{document}

\title{Static Enforcement of Role-Based Access Control}
\author{Asad Ali and Maribel Fern{\'a}ndez
\institute{Department of Informatics, King's College London, Strand WC2R 2LS, UK}
\email{asad.2.ali@kcl.ac.uk}}
\def\authorrunning{A. Ali \& M. Fern{\'a}ndez}
\def\titlerunning{Static Enforcement of RBAC}
\date{}

\maketitle
\begin{abstract}
We propose a new static approach to Role-Based Access Control (RBAC)
policy enforcement. The static approach we advocate includes a new
design methodology, for applications involving RBAC, which integrates
the security requirements into the system's architecture.
We apply this new approach to policies restricting
calls to methods in Java applications. 
We present a language to express RBAC policies on calls to methods in
Java, a set of design patterns which Java programs must adhere to for
the policy to be enforced statically, and a description of the checks
made by our static verifier for static enforcement.
\end{abstract}

\section{Introduction}
The objectives of an access control system are often
	described in terms of protecting system resources against
	inappropriate or undesired user access. 	
 When there is a request for a
	resource, the system must check who triggered the
	request \textit{(authentication)}, check if that user
	has the permission for the request to be fulfilled
	(\textit{authorisation}) and as a result allow or deny the
	request \textit{(enforcement)}. Thus, an implementation of access control
	requires a specification of the rights associated to users in
	relation to resources \textit{(a policy)}. For this, 
        several models of access control have been defined, from
        simple access control lists  giving for each user
        the list of authorised operations, to more abstract models,
        such as the popular Role-Based Access Control (RBAC)
        model~\cite{fer:92:rbac}.
        
Our focus is on enforcement, for which there exist two main
approaches, static and dynamic, with a recently emerged third approach
combining the two: the hybrid approach. The static approach performs
all access checks at compile time, whereas the dynamic approach
performs these at run time.  In short, the static approach enables
policy violations to be detected earlier, facilitating debugging and
reducing the impact on testing, and usually involves a lower run-time
cost. However, the kinds of policies enforceable statically are not as
expressive nor as flexible as those enforceable by the dynamic
approach. We refer to~\cite{hamlen:06:ccem} for a more detailed comparison;
see also \cite{bodden:12:clara} for hybrid analysis of programs, although not
directly applicable to our problem (discussed further in Section
\ref{sec:relw}).

The overall goal of our work is to enforce general access control
policies using a hybrid approach, that is, using a combination of compile-time
and run-time checks. In this paper, we present the first stage of our
work, which is focused solely on static enforcement. 
Our main result is a mechanism to fully verify RBAC policies statically.
More precisely,  we consider implementations of RBAC policies in Java,
where policies restrict method invocations, and present a static source-code 
verifier to enforce the policies. 
Our static verifier ensures that validated  programs  contain no 
unauthorised method invocations. 

RBAC is a widely used policy specification model.
Our static program analysis is applicable to RBAC implementations
under certain important conditions. The first of these is that the source code must
be available at compile time. Secondly, the code should not be modified at run time
through mechanisms such as reflection, therefore our system is aimed at 
non-malicious programmers. Thirdly, the policy should not
change at run time nor should it rely on dynamic information (which
changes throughout execution). The latter condition holds for
the first and second \lq levels\rq\ of the standardised RBAC models: 
\emph{flat-RBAC} and \emph{hierarchical-RBAC} \cite{fer:01:rbacstandard},
if  we disallow administrative changes to the policy 
(Section \ref{sec:eval} discusses in more detail these
restrictions and our plans to relax them in future work).
We therefore provide a policy specification language
which supports resource, permission and role definitions,
and also role hierarchies. 
To the best of our knowledge, these kinds of policies are typically
enforced dynamically in today's available RBAC systems 
(e.g., Java Web Security amongst others~\cite{basin:06:mds,gosling:05:jls}).
One significant reason for this is that during static analysis, it is
difficult to know which regions of code are accessible by users with
which roles. This is because the roles are not usually part of the
application at the design and source level -- they exist only at
run time as part of the (dynamic) security context information. We
have solved this problem through the use of a program design methodology, which
integrates the RBAC model in the system's architecture.

To highlight the problem,
consider the following Java code~\cite{gupta:jee}:\\
\centerline{
\texttt{if(securityContext.isUserInRole(``admin'' )) wipeData();}}

These kinds of code snippets are common in RBAC implementions.
In such cases, a programmer would want to be sure that only the authorised
role (\lq admin\rq , in this example) can invoke the security-critical, or
\textit{protected} method (\lq wipeData\rq , in this example).
This would usually be done using a dynamic check -- the \textit{if} statement
(which in this case utilises Java Servlet API's \emph{isUserInRole()} method
\cite{gupta:jee}),
before any such method invocation. The program would then have to be rigorously
tested to ensure that each role can reach only those invocations that it is allowed
to. It would be reasonable to assume that the number of test cases needed would
increase as the number of roles increases and the number of protected invocations
in the program  increases.
 
Catching errors at an early stage statically aids debugging and
reduces testing time.
 So, since a hierarchical-RBAC  (and also
flat-RBAC) policy is static
 (with administrative changes disabled),
 a program implementing this policy 
should be able to be checked at compile time for policy compliance.
Having said that, just because the policy is static does not mean it
is trivial to statically check the program for policy compliance. Let
us start by removing the dynamic check in our example code -- the
\textit{if} statement -- leaving just the invocation of the protected
method.  We now need to know statically which roles can perform the
invocation.  The difficulty is that the active role exists only in the
security context of the program's execution. In this paper, we show
that it is indeed possible to statically enforce
 hierarchical (or flat)
 RBAC policies.  Moreover, only the assignment of
permissions to roles is assumed to be static, user-role assignments
can change, providing more flexibility.

Summarising, we propose a static solution to RBAC policy enforcement for Java programs 
through the use of new \emph{RBAC MVC} design patterns combined 
with a set of static verification checks made by our static verifier.
The patterns integrate
roles into the program as a set of Model-View-Controller (MVC)~\cite{krasner:88:mvc} 
components (i.e. classes) for each role.
Each role's associated MVC classes act as a
role-specific interface to accessing resources -- protected
methods in resources are invoked in these role classes only.
The flow of the program directs users to the set of role classes
associated to their active role.
Finally, the protected invocations are checked statically for policy
compliance. We present a static verifier, which performs syntactic
checks and call graph analysis to ensure the invocations to methods
belonging to resource classes are made only in role classes,
such invocations are permitted according the policy and role
classes do not invoke methods of components belonging to other roles.

The rest of the paper is organised as follows. After recalling the
basic notions in RBAC and the concept of a design pattern in
Section~\ref{sec:prelim}, we give an overview of the approach in
Section~\ref{sec:conoa}, followed by the definition of the policy
language in Section~\ref{sec:pol}. Section~\ref{sec:patterns}
introduces the RBAC MVC patterns and Section~\ref{sec:staver}
describes the static verifier. The implementation is described in
Section~\ref{sec:impl}. Section~\ref{sec:relw} discusses related work
and Section~\ref{sec:conc} concludes and discusses future work.

\section{Preliminaries}
\label{sec:prelim}\label{sec:jee}

\emph{Role-Based Access Control} is a mechanism to protect 
	resources from unauthorised use in an organisation, where
	instead of specifying all the accesses each user is
	allowed to execute, access authorisations on objects
	are specified for roles \cite{fer:92:rbac}. Each role is 
	given a set of access rights, and each user is given
	a set of roles so
	that only authenticated users who have activated
	the required role can 
	access and use the restricted resources. Roles can
	be arranged in a hierarchy, where a more senior role
	\lq subsumes\rq\ another; the senior role
	inherits the permissions of the subsumed role
	and can be assigned further permissions.

A \textit{design pattern} describes a  particular recurring design problem that
arises in a specific design context, and presents a generic scheme for
its solution \cite{buschmann:96:posa}. Patterns are usually described using
the semi-formal Unified Modelling Language (UML) notation, showing
its constituent components, their
responsibilities and relationships, and the way in which
they collaborate.
The goal of patterns is to provide a
mechanism to guide the implementation of a solution to a specific problem.

 Our work utilises concepts from the well known and widely used
\emph{Model-View-Controller (MVC) pattern}. This pattern achieves 
separation of concern for user interaction~\cite{krasner:88:mvc},
separating data processing from user interaction, allowing
both to be modified independently. Data processing is handled
by \emph{model} components, data presentation and user-interaction are
handled by \emph{view} components and the communication between
these two is handled by \emph{controller} components.

\section{Conceptual Overview of Approach}
\label{sec:conoa}
\begin{figure}[ht]
	\centering
	\includegraphics[width=0.7\textwidth]{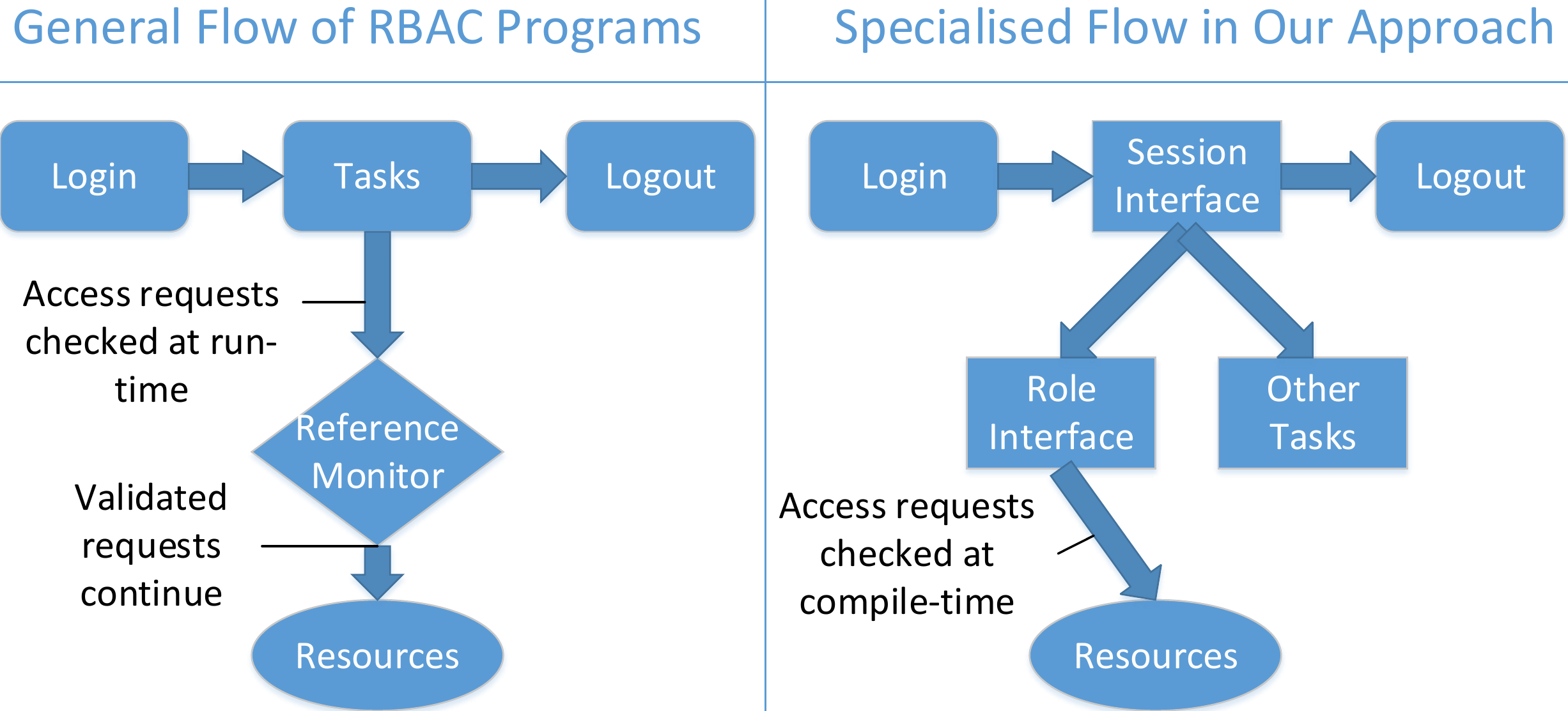}
	\caption{General and specialised flow of programs that enforce RBAC}
	\label{fig:generalspecialflow}
	\end{figure}

Programs that restrict access to resources from users
typically involve an initial user authentication phase, where users
log in and retrieve their access rights, then allowing users to
undertake user tasks which may involve accessing resources, and
finally logging out of the system. We present a simplified model of
the general flow of a program which implements RBAC in the left-hand
side of Figure \ref{fig:generalspecialflow}.
In RBAC, authentication also involves retrieving and activating the
role(s) associated to the user, and logging out also involves
deactivating the role(s). Controlling access most commonly takes place
 between \lq Tasks\rq\ and \lq Resources\rq , for example through a
reference monitor intercepting all access requests made to resources
at run time, stopping those requests which are unauthorised.

In our approach, we divide user tasks into three groups: \emph{role tasks}
that users with certain roles in the system can
perform (these may access resources), \emph{other tasks} that the policy
is not directly concerned with and \emph{session tasks} related to the
functioning of the session.
After a successful log-in, users are presented
with a \emph{session interface}. This is made up of MVC components
that implement session tasks e.g. log-out. The session interface
will retrieve and hold a list
of all the roles assigned to the user in the policy. From here, the
user can choose to perform role tasks by selecting one of the retrieved
roles, resulting in the session interface displaying 
the \emph{role interface} for the selected role. Each role has an associated
role interface which implements its role tasks through a set of
MVC components: a set of view components, one controller and one model
component. Direct access to resources is prevented; resources
can only be accessed through a role interface. The user interacts
with a \emph{Role View} which communicates
with its \emph{Role Controller}, which communicates
with its \emph{Role Model} which finally may access a resource.
In this way, if an access request is found at compile time
in a class that is part of a role interface, then the role
that the interface belongs to is the role that can reach
and execute this request. Our \emph{RBAC MVC} patterns guide the
implementation of the program to achieve this flow.

We now define the core concepts in our approach.

	\begin{mydef}[Resource]
		 A \textit{resource} 
		 is realised
		as a \emph{resource class}	containing some methods whose invocation
		needs to be restricted. Invocations are restricted for instances
		of resource classes.
	\end{mydef}
%
Methods in resource classes  are categorised as follows.
	
\begin{mydef}[Actions and Auxiliary Methods]
An \emph{action} is a method in a resource class that must only be invoked by those users
 with the permission to do so.
An \emph{auxiliary} method is a method in a resource class that is not part of
 the policy definition. Such methods are usually required for the correct
initialisation and operation of a class, and should  not be invoked
directly by users.
\end{mydef}


	\begin{mydef}[Permission]
	\label{mydef:perm}  
	A \emph{permission} is a pair $[res, act]$ where $res$ is the name of a resource
	and $act$ is the name of an action of that resource. The
        action is allowed to be invoked on any instance of that resource
        class by the role (see Definition
        \ref{mydef:role}) which the permission is assigned to.
	\end{mydef}

	\begin{mydef}[Access]
	An \emph{access to a resource} is an invocation or call to an action
	method of an instance of a resource class.
	\end{mydef}
	
	\begin{mydef}[Task]
	\label{mydef:task}
	We divide the concept of a user task into three groups as follows. Firstly,
	a \emph{role task} is an operation, or business function, to be performed
	by an authorised user in a specific role, which could involve the invocation
	of one or more actions on resources. Secondly, a \emph{session task} is an
	operation required to correctly manage the session e.g. log-in and log-out.
	Thirdly, an \emph{other task} is an operation or function that is executable
	by all users, regardless of the notion of role as it does not access resources
	(in the access control sense).
	\end{mydef}

	An example of a role task is as follows. A user in an Admin role in a GP surgery
        may need to perform the task \emph{\texttt{registerPatient}},
        which would involve a call to an action
        e.g. \emph{\texttt{addPatient}} in the
        \emph{\texttt{Patients}} resource.
 
	\begin{mydef}[Role]
	\label{mydef:role}
	 At the policy level, a \textit{role} consists of a name and a
         list of permissions to access resources.
     \end{mydef}
 
In the context of our system, a role is implemented by a set of MVC
components: a set of Role View components (i.e. classes), a Role
Controller class and a Role Model class (as defined below).  Together,
these provide a \emph{role-specific interface} for the user to perform
tasks. We define these components below.

\begin{mydef}[Role Model]
	\label{mydef:rolemodel}
	 A \textit{Role Model} provides role-task methods which should only call
	those actions that are permitted for its role. Its name must be 
	prefixed with the name of the role, followed by \lq Model\rq .
	\end{mydef}
	
	\begin{mydef}[Role Controller]
	\label{mydef:rolecontroller}
	 A \textit{Role Controller} acts as an intermediary between the Role Model
	and View classes. Its name must be 
	prefixed with the name of the role, followed by \lq Controller\rq . 
	Role Controller methods are invoked in 
	Role View classes to communicate with the Controller.
	\end{mydef}
	
	\begin{mydef}[Role View]
	\label{mydef:roleview}
	 A \textit{Role View}  provides (part of) the user-interface for users
	 to execute the tasks of their role. Its name must be 
	prefixed with the name of the role, followed by \lq View\rq\ (followed
	by any valid Java identifier).
	\end{mydef}

	For any role \textit{r}, its single associated role model class contains
	the code that performs the tasks \textit{r} can do
	in the system. The role's multiple associated role view classes and its
	single associated role controller class, provide the means for  users
	that have activated this role to access these tasks (and perform them).
	
		An example of a set of role components is as follows.
        Firstly, a role model class \emph{AdminModel}, which
       	provides a role task \emph{\texttt{registerPatient()}} that calls
       	on the \emph{\texttt{addPatient()}} action in the
       	\emph{\texttt{Patients}} resource. Secondly, a set of role
        view components \emph{AdminViewPatients,
        AdminViewAppointments,} e.t.c. Thirdly, a role
        controller \emph{AdminController} acting as an
       	intermediary between the role view and model components.
       	
   \begin{mydef}[Session]
    \label{def:session}
    A \emph{session} is the state of the program in which an authenticated
    user is able to perform the three kinds of  tasks in the system.
    The session has a
  	user interface composed of a session-specific interface,
  	the role-specific interface (made up of Role MVC components discussed
  	above) of the current active role and any interfaces implementing
  	other tasks. The session-specific interface
  	is made up of a set of MVC components: one Session Model,
  	one Session Controller and a set of Session View classes.
  	The Session Model implements the
  	session tasks which are: log-in/authentication,
  	role activation, log-out, calling a role-interface and
  	calling classes that implement other tasks. The Session
  	Views and Controller provide the means for the user to access
  	these session tasks. The session-specific
  	interface is always active so that the session tasks are
  	always available to the user. 
  	We, of course, have minimum expectations such as
  	log-out only being available if logged-in and so forth.  The
  	session-specific interface also allows the user to interact
  	with the system via their role by calling a role interface,
  	or without their role thus calling other-task implementing classes.
  	Names of session classes start with the string \lq Session\rq\ followed
  	by either \lq Model\rq , \lq Controller\rq\ or \lq View\rq . For the latter, since
  	there can be many Session View classes, any valid class identifier
  	(in Java) is allowed to follow in the name.
   	\end{mydef}
   	 The classes required for
  	the session -- Session Classes -- constitute part of our Trusted
  	Computing Base (TCB); the other part is the actions, which we trust
  	behave safely. The session classes should contain the minimum
  	code necessary to implement session tasks, so that the TCB 
  	is small. We perform few checks, and exercise few constraints,
  	on session classes in order for their implementation to be as
  	flexible as possible. Therefore, we do not deal with authentication
  	in this paper. However, an important aspect of an RBAC system is
  	the \emph{active role}, which is to be implemented by the session classes.
  	We give guidelines for implementing the active role below
  	by first defining the concept of \emph{authenticated user}.
  	
  	\begin{mydef}[Authenticated User]
  	\label{def:authUser}
  	An \emph{authenticated user} is a user that passes the authentication
  	process, which is left open and unrestricted for the programmer in our
  	approach except for one condition: after successful authentication,
  	the session model class contains a list \emph{retrievedRoles} containing
  	the names of the roles given to that user in the policy.
  	\end{mydef}

	\begin{mydef}[Active Role]
	\label{def:activation}
	The \emph{active role} is the single role $r_a$ selected by the user from
	the \emph{retrievedRoles} (see Definition \ref{def:authUser}) whose role
	interface is being displayed to the user so that they may perform
	the role tasks associated with $r_a$.
	\emph{Role activation} constitutes storing  $r_a$ in a field called
   	\emph{activeRole} in the session model class
   	and invoking the role controller of that role. This
   	process is achieved in a method \emph{activateRole()} in the
   	session model class (See Definition \ref{def:session}). This will result
   	in presenting a  
        role view of the active role's
   	role interface to the user by composing it with the session
   	interface.
	\end{mydef}


\section{Policy Language: JPol}
\label{sec:pol}

We define a policy specification language for hierarchical-RBAC, called JPol,
 where resources,
        together  with their associated lists of actions, and roles,
        together  with their associated permissions, can be declared.
        To simplify, we assume that only the access requests that are
        allowed are expressed, so all other requests are not allowed.
        The
        policy file will be parsed and represented as a set of
        tables, to be used only at compile-time by the
        static verifier in order to perform the
        access checks.
	
	The policy language does not support user definitions and
        user-role assignments, since we do not deal with
        authentication in this paper. 
	Since with our static verifier, only the roles which have been declared in
	the policy will	be permitted to be assigned to users, the resources
	will still be protected because each role will have been checked at
	compile-time to ensure it does not perform any illegal access
	requests. The proposed approach is flexible:
	new users can be added to the system and
	user-role assignments can change depending on changes within the
	organisation.
	
\subsection{Syntax and Representation}
	\label{sec:polAS}
	The policy language adopts an
        object-oriented, Java-like syntax designed to make the policy
        implementer's transition from target program language, Java,
        to policy language as effortless as possible.  However, as we
        will see later, the static verifier relies solely on the
        information generated as a result of parsing the policy
    	file. Thus, the syntax of the policy language can change
        and be adapted to any environment using hierarchical-
        (or flat-) RBAC. We could, for instance, use one of the
        existing RBAC specification languages.
	
\begin{figure}[h!]
	\centering
	\includegraphics[width=0.8\textwidth]{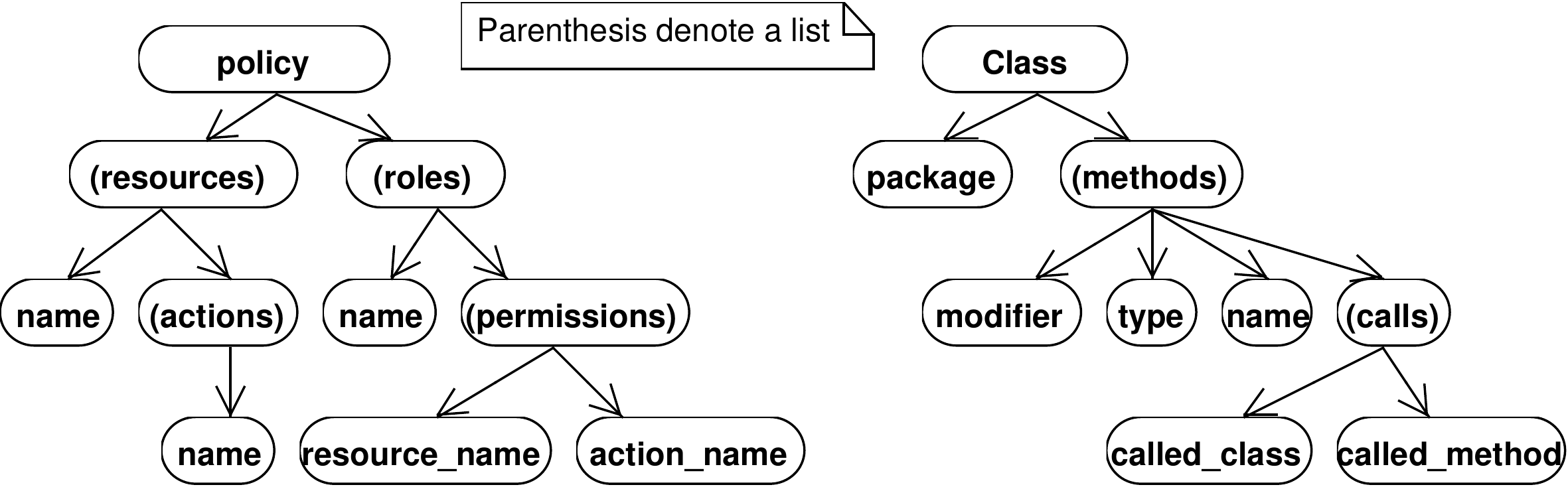}
	\caption{Abstract syntax of policy on the left and a class on the right}
	\label{fig:polClassAS}
	\end{figure}

	The grammar of the policy language  is
	as follows, where the keyword \emph{subsumes}
	indicates role inheritance. The abstract syntax of the
	policy language is illustrated in Figure \ref{fig:polClassAS}.
	

\lstset{
  basicstyle=\footnotesize
}

	\begin{lstlisting}
stmts          ::=   (stmt `;')+
stmt           ::=   decRole | decRoleSubsume | decRes 
	             | addActRes | addPermRole
decRole        ::=   `Role' ID `=' `new' `Role' name	
decRoleSubsume ::=   `Role' ID `=' `new' `Role' name
	             `subsumes' ID	
decRes         ::=   `Resource' ID `=' `new' `Resource' name	
addActRes      ::=   ID `.' `addAction' name
addPermRole    ::=   ID `.' `addPermission' permission
name           ::=   `(' ID `)'
permission     ::=   `(' ID `,' ID `)'
\end{lstlisting}

The parser for the policy specification language checks that a policy
declaration is syntactically correct, producing the Abstract Syntax
Tree (AST) shown in Figure \ref{fig:polClassAS}.  It then generates
intermediate data structures -- tables called \lq Resources\rq\ and \lq Roles\rq\
containing the information  needed for the static
verifier.

\begin{figure}[h!]
	\centering
	\includegraphics[width=0.70\textwidth]{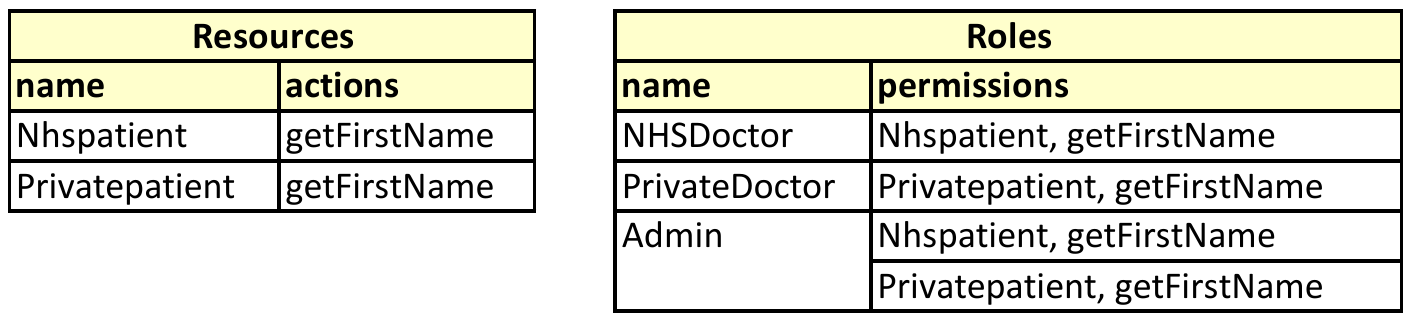}
	\caption{Example Roles and Resources Tables Representation}
	\label{fig:resroles}
	\end{figure}

Listing \ref{lst:resroles} shows an example specification in JPol for
patient-related resources and permissions for roles in an example
GP/doctor's surgery, with the resulting
tables \lq Resources\rq\ and \lq Roles\rq\ shown in Figure \ref{fig:resroles}.

\lstset{basicstyle=\footnotesize\ttfamily, frame=single,
  caption=Example JPol code declaring Resources with their actions
  and Roles with their permissions,
  label=lst:resroles}

    \begin{lstlisting}
Resource nhspatient = new Resource(`Nhspatient');
nhspatient.addAction(`getFirstName');
Resource privatepatient = new Resource(`Privatepatient');
privatepatient.addAction(`getFirstName');
Role nhsdoctor = new Role(`NHSDoctor');
nhsdoctor.addPermission(`Nhspatient', `getFirstName');
Role privatedoctor = new Role(`PrivateDoctor');
privatedoctor.addPermission(`Privatepatient', `getFirstName');
Role admin = new Role(`Admin');
admin.addPermission(`Nhspatient', `getFirstName');
admin.addPermission(`Privatepatient', `getFirstName');
    \end{lstlisting}
    
\subsection{Semantics}
 We can state the semantics of
 the policy language in a concise  manner by mapping the abstract syntax
to elements of the RBAC model: there is a one-to-one correspondence between
the resources, roles and permissions specified in JPol and in the RBAC model. 
In particular, an \lq addPermission\rq\ statement in JPol syntax (see the grammar 
rule for \lq addPermRole\rq\ above) corresponds directly to a permission in the
RBAC sense. Therefore, we can define policy satisfaction as follows.
 
\begin{mydef}[Policy Satisfaction]
\label{def:semantics} 
A Java program satisfies a JPol policy  if, for any invocation
 $res.m$ that exists in the program,     
         where $res$ is an instance of a resource class $Res$ and $m$ an action,
         only authenticated users with active role $r$, such that the
 JPol policy specifies the permission $[Res, m]$ for $r$,  can perform $res.m$.
\end{mydef}

\section{Program Design Patterns - RBAC MVC}
\label{sec:patterns}
In order for the target program to be statically checked for policy compliance, it
must follow our RBAC MVC Patterns described below. 



\subsection{RBAC Model, Controller, View  and Session Patterns}
\begin{figure}[h!]
	\centering
	\includegraphics[width=1.0\textwidth]{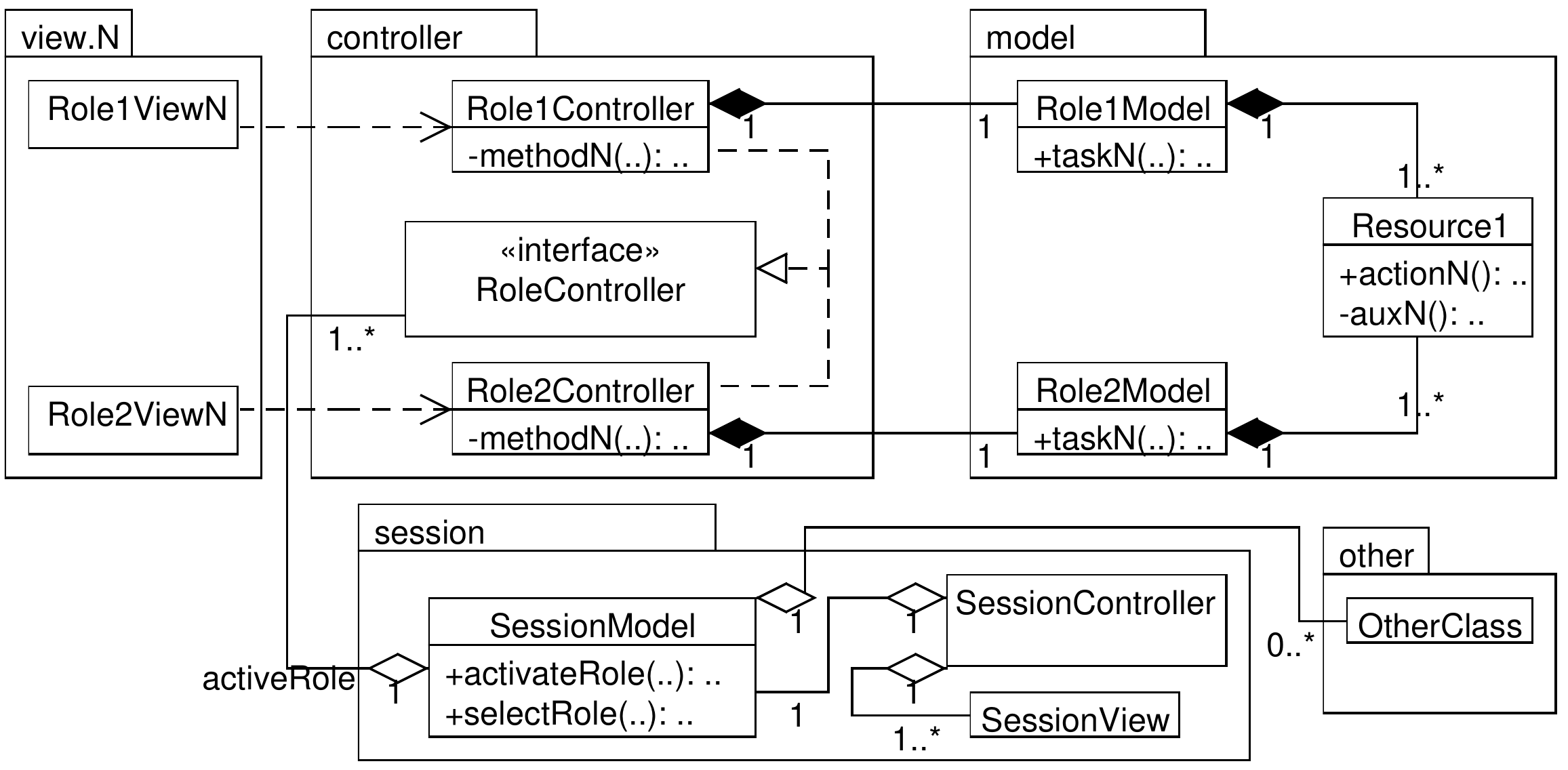}
	\caption{UML Class Diagram of RBAC Model, RBAC Controller, RBAC View and RBAC Session patterns.
	Note that \emph{N} represents any valid identifier in Java.}
	\label{fig:rbacmvc}
	\end{figure}

The class diagrams of the patterns are shown together in Fig. \ref{fig:rbacmvc}. RBAC Model
contains only packages with names containing \lq model\rq , describing the design
of resource and role model classes.
RBAC Controller adds packages with names containing \lq controller\rq ,  describing the design
of role controller classes. The empty interface class \lq RoleController\rq\ simply groups
all role controllers to simplify the link with session classes.
RBAC View adds packages with names containing \lq view.n\rq\ (where \emph{n} represents any
valid package identifier in Java) to
these, describing the design of sets of role view classes.
RBAC Session adds the
package \lq session\rq , to guide the implementation
of two key RBAC concepts: activating a role and users having multiple roles
being able to switch between them. It also adds the package 
\lq other\rq\ containing other classes, linking the session classes to them.

\section{Static Verification}
\label{sec:staver}
Our source-level static verifier takes as input a well-formed program, 
which is defined as follows:

\begin{mydef}[Well-formed program]
\label{def:wellformed}
A well-formed program consists of a (syntactically correct) JPol
policy file and a Java program that implements the RBAC MVC patterns
defined in Section~\ref{sec:patterns}.  Implementing the patterns
means: there is a set of session classes (one model, one controller
and multiple views associated to the session), a set of resource
classes, a set of role classes (sets of one role model, one role
controller and multiple role view classes) and a set of classes which
do not fit into the other groups.  In particular for session classes,
they: correctly authenticate users, activate the correct role(s)
allowed for the user, switch roles correctly for the retrieved and
selected roles.
\end{mydef}

A well-formed program might contain unauthorised calls to actions on resources. 
The static verifier should reject a program if an access violation is
found, else accept it. In other words, it should only accept programs
that satisfy the policy (see Definition \ref{def:semantics}). In this
section, we describe high-level details of the  static
verifier, which is composed of a parser that 
generates abstract syntax representations of
the policy and program, and populates tables,
and a checker that
uses the abstract syntax tree and tables to check that the program
satisfies the policy.
\subsection{Parsing}
\label{sec:parser}
Section \ref{sec:polAS} described the 
AST and tables generated by parsing the policy. We now
describe the process of parsing the program.
%

The parser generates an AST for each class using standard parsing rules 
for Java.
Figure~\ref{fig:polClassAS} (right) shows a simplified AST of a class; 
note that for the node \emph{called_class}, if it is an object, our parser
resolves the object to obtain the name of its class. In addition,
the parser groups classes into Resource classes, Role Model classes,
Role Controller classes, Role View classes, 
Session classes and Other classes (the latter
are any that do not fit into other groups).


In order to group each class, we use naming restrictions on the
package and class names, described informally as follows.
Names of resource classes
must be the same as the name of a resource given in the policy,
names of session classes must begin with string \lq Session\rq\ and can
then be followed by any valid identifier in Java - this applies to
session model, session view and session controller classes
which are grouped together into one group.
Names of role model and role controller classes must begin with
the name of a role given in the policy followed by the string
\lq Model\rq\ and \lq Controller\rq\ respectively. Names of role view classes
must begin with the name of the role followed by the string \lq View\rq\
and can then be followed by any valid identifier in Java. 


This phase also generates
tables containing the names of all classes
in each group except \emph{Other classes}. We call these tables
\emph{ResourceClasses}, \emph{RoleModelClasses}, \emph{RoleControllerClasses},
\emph{RoleViewClasses} and \emph{SessionClasses}.
This is to simplify the process of looking up called classes in the checks made
by the verifier (discussed below).

\subsection{Static Verifier Checks}
\label{sec:implchecks}
%
The checks performed on each group of classes are described informally at a
high level as follows.

\subsubsection{Resource Class Checks}
\label{sec:rescc}
The checks on resource classes, performed by a subprogram of the verifier
called \emph{ResourceClassChecks}, are described below. 

\begin{enumerate}
\item For each method (each element of the node \emph{methods},
see Figure~\ref{fig:polClassAS}),
we search the actions sub-table (generated when parsing the policy, 
see Figures~\ref{fig:polClassAS} and \ref{fig:resroles})
for \emph{Class.name} (which is the name of a resource) then:
	\begin{enumerate}
	\item If the method name is in this sub-table, then the
		value of the node \emph{modifier}
		(see Figure~\ref{fig:polClassAS}) must be \lq public\rq .
	\item Else the value of the node \emph{modifier} must be
		\lq private\rq .
	\end{enumerate}
\item For each call (each element of the node \emph{calls}, see Figure~\ref{fig:polClassAS})
	we check that:
	\begin{enumerate}
	\item \label{check:notRoModel} The called class 
	(the node \emph{called_class}) is not the name of
	a role model class. This is done by searching the names
	of classes in the table \emph{RoleModelClasses} generated
	by the parser.
	\item \label{check:notRoContr} The called class is not the
	name of a role controller class. This is done by
	searching the names of classes in the table
	\emph{RoleControllerClasses} generated by the parser.
	\item \label{check:notRoView} The called class is not the
	name of a role view class. This is done by searching
	the names of classes in the table \emph{RoleViewClasses}
	generated by the parser.
	\item \label{check:notSession} The called class is not the
	name of a session class. This is done by searching
	the names of classes in the table \emph{SessionClasses}
	\end{enumerate}
\end{enumerate}

\subsubsection{Role Model Class Checks}
\label{sec:romcc}
The checks on role model classes, performed by a subprogram 
called \emph{RoleModelClassChecks}, are described below. The
role name the class belongs to is obtained by removing the
substring \lq Model\rq\ from \emph{Class.name}.
\begin{enumerate}
\item For each call, we check that:
	\begin{enumerate}
	\item \label{check:perm} If the called class is a resource
	class, then
		\begin{enumerate}
		\item If the called method
		(the node \emph{called_method}) is an action,
		which is done by searching the \emph{actions} sub-table
		for that resource in the table \emph{Resources}
		generated when parsing the policy, see 
		Figures~\ref{fig:polClassAS} and \ref{fig:resroles},
		then the pair of values
		[\emph{called_class}, \emph{called_method}]
		must appear in the permissions for the associated role
		of the	class (done by searching the \emph{permissions}
		sub-table of the matching role in table \emph{Roles})
		\end{enumerate}
	\item The called class is not the
	name of a different role model class.
	This is done by checking if the called class
	contains the substring \lq Model\rq , then the
	name of the class must be the same as
	the value in the node \emph{Class.name}.
	\item The subsequent checks are equivalent to Section \ref{sec:rescc}
	Checks \ref{check:notRoContr}, \ref{check:notRoView} and \ref{check:notSession}.
	\end{enumerate}
\end{enumerate}

\subsubsection{Role Controller Class Checks}
\label{sec:roccc}
The checks on role controller classes, performed by a subprogram 
called \emph{RoleControllerClassChecks}, are described below. The
role name that the class belongs to is obtained by removing the
substring \lq Controller\rq\ from \emph{Class.name}.
\begin{enumerate}

\item For each call 
	 we check:
	\begin{enumerate}
	\item If the method called is an action, we do the same check as 
        Section \ref{sec:romcc}
	Check \ref{check:perm}.
	\item \label{check:diffRoModel} The called class is not the name of a
	different role's role model class. This is done
	by checking if the value of \emph{called_class}
	contains the substring \lq Model\rq , then
    remove this substring and check that this \emph{called role}
    matches the role that this role controller class belongs to.
	\item \label{check:diffRoContr} The called class is not the name of a
	different role's role controller class. 
	This is done by checking if the value of
	\emph{called_class}	contains the substring \lq Controller\rq , 
   	then this value must be the same as the
   	value in \emph{Class.name}.
	\item \label{check:diffRoView}
	The called class is not the name of a
	different role's role view class.
	This is done by finding the suffix of the value
	in \emph{called_class}
	which begins with the string \lq View\rq\ and ends 
	at the end of the value, then removing this entire suffix.
	The remaining value must match the role name of this
	role view class.
	\item The called class is not the name of a session class.
	\end{enumerate}

\end{enumerate}

\subsubsection{Role View Class Checks}
\label{sec:rovcc}
The checks on role view classes, performed by a subprogram 
called \emph{RoleViewClassChecks}, are described below. 
The role name the class belongs to is obtained by finding
the suffix of the value in \emph{Class.name} which begins
with the string \lq View\rq\ and ends at the end of the value,
then removing this entire suffix.
\begin{enumerate}
\item For each call we check:
	\begin{enumerate}
	\item If the method called is an action, we do the same check as Section \ref{sec:romcc}
	Check \ref{check:perm}.
	\item The remaining checks are equivalent to 
	Section \ref{sec:rescc} Checks \ref{check:notRoModel}
	and \ref{check:notSession}
	and Section \ref{sec:roccc} Checks \ref{check:diffRoContr}
	and \ref{check:diffRoView}.
	\end{enumerate}

\end{enumerate}

\subsubsection{Session Class and Other Class Checks}
For each method invocation within a session class, we check that 
the method called does not belong to a resource class or role model class
(due to Definitions \ref{def:session} and \ref{def:activation}).
For each method invocation in an other class, we check that it is not 
calling a method belonging to resource classes, or to role or
session classes.
We omit details of these checks due to space restrictions.

\subsection{Properties}

The static verification checks described in the previous sections
ensure that programs that pass the checks do not perform invalid
access requests. More precisely, the source code of programs satisfies
the propositions stated below, for which we first define the notion
of \emph{OK-program}.

\begin{mydef}[OK-program]
\label{def:ok}
A program $P$ is OK, written $OK(P)$, if its actions
are \lq public\rq\ and auxiliary methods are \lq private\rq ; resource classes
do not invoke methods of a role model, role controller, role view
or session class; role model methods do not invoke session classes,
role controller classes, role view classes or
an action that is not allowed by the policy for the associated role;
role controller classes
do not invoke session classes 
or an action that is not
allowed by the policy for the associated role;
role view methods
do not invoke role model or session classes or an action that is not
allowed by the policy for the associated role;
role classes do not call classes belonging to other roles;
session classes do not invoke resource classes or role classes
except for role controllers
and role views;
other class methods do not call role, resource, or session classes.
\end{mydef}

\begin{proposition}
\label{the:accept}
Let $P$ be a well-formed program. $P$ is
accepted by the static verifier if and only if $OK(P)$.
\end{proposition}

\begin{proposition}
A well-formed program $P$ accepted by the verifier
satisfies the policy (see Definition \ref{def:semantics}).
\end{proposition}

We provide an intuitive explanation of the propositions as follows.
According to
 Definition \ref{def:semantics} we need to show that only
 authorised users with active role $r$ having permission
 $[Res,m]$ can invoke the action $m$ of an instance of $Res$.
 Let $res.m$ be a call to $m$ in the program $P$, for which
 the parser has identified the called class to be $Res$ and
 the called method to be an action $m$.  Since
 $P$ is \emph{well-formed}, by Definition~\ref{def:wellformed}
 it implements the RBAC MVC patterns. Therefore, the user $u$ 
executing $res.m$  has been authenticated
 and is in a session, where by Definition \ref{def:session},
 one of $u$'s roles, say $r$, has been activated.

 By Definition \ref{def:activation}, this implies that $r$'s role
 controller has been invoked. Moreover, since $P$ has been accepted by
 the verifier, by Proposition~\ref{the:accept}, $OK(P)$.  Once the role
 controller for $r$ has been invoked, by Definition~\ref{def:ok} the
 Java code executed from the role classes associated to $r$ contains
 only invocations $res.m$ that are authorised by the policy, there are
 no calls to methods in role classes belonging to other roles,
 and any call to a method in
 a class which is not one of $r$'s role classes (except for a call to a session class)
will not contain invocations to
 actions or to role classes. Note that the only classes outside role classes
 which could call a role class are session classes, which, since the
 program is well-formed, must satisfy the requirements of the RBAC-MVC pattern.
 In particular, we trust the calls to role classes made in session classes.

To provide flexibility to programmers, we have allowed actions to be freely invoked
within resource classes. 
We assume that actions are not restricted in their behaviour  (i.e.,
the policy specifies the actions that a role is allowed to call, and it does not
restrict the invocations within those actions).
The session classes are the critical part of the program in our approach,
in which role class invocations are trusted and not verified. The minimal Trusted Computing
Base in our approach is therefore the action methods and the session classes.
 In future work, we will extend the verifier to include checks within actions, to alert
programmers if an action calls another action not allowed by the policy. 

\section{Implementation and Evaluation}
\label{sec:impl}
\paragraph{Implementation} 
Our implementation consists of a JPol policy parser, produced using
the ANTLRWorks tool~\cite{bovet:08:antlrworks}, and a static analysis program
which are both part of a plug-in we have produced for the Eclipse
Integrated Development Environment (IDE) \cite{eclipse}
\footnote{The plugin and a sample program can be found at 
www.inf.kcl.ac.uk/pg/aliasad}.
Eclipse plugins are able to use the
Java Development Tools (JDT) Application Programming Interface (API)
provided by Eclipse, whose benefits include simplifying static code
analysis. In Java, there are three ways to 
invoke a method; either invoking a (\lq static\rq ) method on a class e.g. 
\lq ClassName.methodOne()\rq , invoking a method on a variable e.g.
\lq x.methodOne()\rq\ or invoking a method on the object returned by another
method call e.g. \lq x.methodOne().methodTwo()\rq . 
Using JDT we can get the type binding for variables and method invocation
expressions, and so we can check if a resource's actions are being called
or if one role's components invoke another role's components. This is
sufficient to implement all the static checks discussed in Section
\ref{sec:staver}. 
We have tested our plug-in on a simple doctor's surgery web database
application implemented in Java Enterprise Edition (JEE) (refer to
\cite{gupta:jee} for an overview of JEE).
The tool outputs helpful error messages
in Eclipse's editor window, consisting of the class name and line number
where the error occurs, the kind of error that has occurred (e.g.
\lq Invocation not permitted\rq ) and a description of why that error could
have occurred. 

\begin{figure}[ht]
	\centering
	\includegraphics[width=0.8\textwidth]{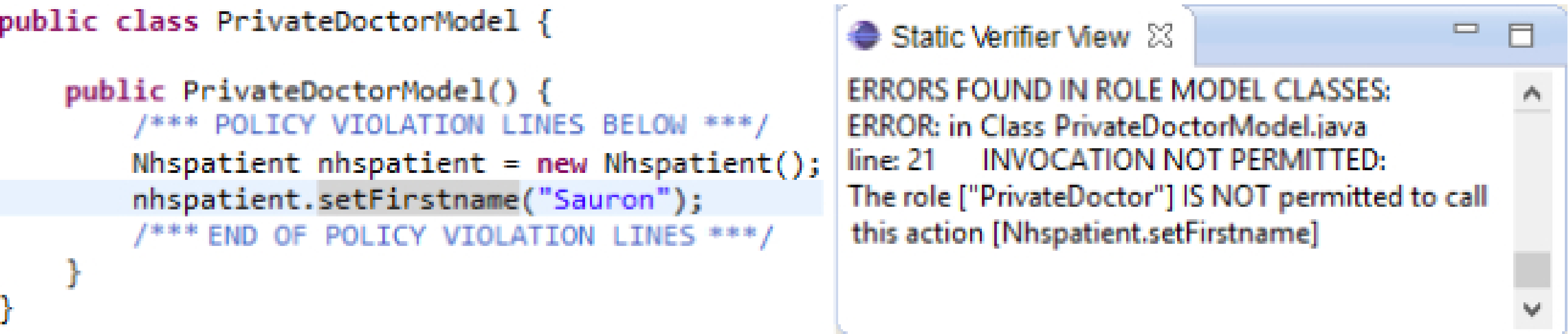}
	\caption{An example error caught by the static verifier}
	\label{fig:invNotPermitted}
	\end{figure}

See Figure \ref{fig:invNotPermitted} for an example
of the verifier catching an error, utilising the policy fragment
described in Listing \ref{lst:resroles}.

\paragraph{Evaluation}
\label{sec:eval}
The static approach to access control enforcement has limitations.
Firstly, the policy cannot change after
compilation, which prevents administrative changes in policies such as
changing role permissions.
Secondly, permissions
cannot be based on any information that changes at run-time. For flat or
hierarchical RBAC policies, this is sufficient. 
For more general versions
of RBAC, these two restrictions can be relaxed by combining the static approach
with a dynamic one, 
which will be the subject
of future work.  User-to-role assignments
can change in our approach and these tend to change more often than
permission-to-role assignments. Thirdly, the program code must be
available at compile-time
in order to analyse it statically.



As a result of the analysis,
policy enforcement is done exclusively at compile-time, aiding debugging since
the program requires no run-time checks for policy enforcement.

The limitations of the static approach do not mean that it is not useful.
A policy commonly contains some static  and some dynamic parts
(even though these may not be clearly separated). For example,
just after log-in, usually
some static
part of the policy will be in effect.
Therefore,  the static
approach can be used in combination with dynamic checks  within a
hybrid checker. 

Using our design pattern, it takes an initial effort
for an architect/programmer to design/implement an initial set of resource classes
and one set of Role MVC classes. 
After this initial stage, designing/implementing the program becomes easier
than without using the patterns. Our patterns help to relate the functionality
of the program with the roles that can access that functionality. Adding this
related functionality becomes easier - achieved just by adding more sets of
Role MVC classes.
Our pattern also helps in the design of resources because
it helps to clearly separate the resources from the rest of the program.
Current limitations of our patterns are that roles that have many
similar operations will require completely separate Role MVC classes,
possibly duplicating code. Moreover, role hierarchies are declared in the
policy but not reflected in
the design of the program; the permissions of a subordinate role are copied
to the senior one and this data is used in the static analysis only.
Reflecting role hierarchies in the program
would reduce code duplication, which we intend to address in future work. 

Lastly, it is difficult to compare the performance of a program
designed using a pattern and the same program designed without using
it. Performance is not usually taken into consideration when designing a 
pattern, especially in the case where performance gains are not the main goal of
a pattern - as in our approach. We can be sure
that in our approach, policy enforcement will have no impact on run-time
resources, since no access-checks will be made at run-time. 

\section{Related Work}
\label{sec:relw}

Formal approaches for the verification of properties of access control
policies usually rely on purpose-built logics or rewrite-based
techniques~\cite{BonattiS03,SohrK:Isabelle,SantanaA:phd,BertolissiC:ppdp}.
In this paper, we have focused on verifying that a program enforces a
policy, rather than on proving properties of the policies. Bodden
et al. \cite{bodden:12:clara} enforce security properties in programs using
a hybrid approach. They generate code for run-time checks, then
perform compile-time analysis to eliminate some of these.
In their approach, the access control
enforcement of (static) roles would not be possible at
compile-time, because they cannot determine, at compile-time,
the access requests that each role can make. Our design pattern solves
this. Therefore, in their approach, a static RBAC policy would be
enforced dynamically.

In Java Web security, there are two types of access control
using roles: declarative and programmatic security. Both of these
use dynamic checks to restrict access to methods,
the former uses XML based permission declarations whilst the latter
uses provided Java methods such as \texttt{isUserInRole()} \cite{gosling:05:jls},
\cite{basin:06:mds}. Our
approach requires no dynamic checks.\\
\indent In design-level security, generally, security restrictions
are specified at the design stage of a program's lifecycle. An example is
Model Driven Security~\cite{basin:06:mds}. 
In this work, code to perform access checks is generated from access control
specified in the UML model of the application. The security code generated
for Java utilises Java Web security mechanisms which are dynamic (as discussed above),
whereas our approach uses static checks.\\
\indent There exists a body of work on security patterns. In \cite{priebe:04:pattern},
the authors describe access control, specifically RBAC and
Metadata-based Access Control, using patterns and run time
checks. Steel, Nagappan and
Lai \cite{steel:06:csp} propose several security patterns that are
specifically targeted towards securing Java (Enterprise) applications. 
Their work takes a dynamic approach to enforcement; in fact, all patterns that
we have seen that aid in the implementation of a policy rely on
dynamic mechanisms. Many 
patterns in \cite{steel:06:csp} can and should be used in conjunction with our patterns to
secure the overall application aside from enforcing RBAC statically.\\
\indent Zarnett et al. \cite{zarnett:10:annot} enforce RBAC in Java using proxy
objects in Remote Method Invocation. Their work has the effect of removing
the need for
run-time access control checks, however their approach relies on
annotations. Understanding where a specific annotation should go can
be a difficult task, especially in large programs. Moreover,
specifying the policy via annotations leaves the policy fragmented
throughout the program. In our approach, we check that the policy has
been implemented correctly, e.g. that all the resources and roles have
been implemented, however they have no such verification techniques
since there exists no central policy specification, which means that
errors are discovered later (at run time). Also, recalling the model
in Figure \ref{fig:generalspecialflow}, their approach enforces access
restrictions at the level of the \lq Resources\rq , by creating proxy
objects containing only those methods which are authorised for the
currently active role. Our approach enforces access control at the
level of \lq Tasks\rq , where instead of creating proxy objects of each
resource for each role, all authorised methods for each role are
provided by a role-specific user-interface. 

\section{Conclusion and Future Work}
\label{sec:conc}
We have described a new system to statically check that a target
program respects its RBAC policy. 
If the program
successfully passes the static verifier's checks, then when using the program,
the logged in user can only call those methods that have been authorised for the
role currently activated for them. Therefore, no run-time access checks are needed. 

In future work, 
we will develop a hybrid approach for policies with  dynamic conditions, 
inlining code in the program to check these at run-time. This
hybrid approach would utilise our concept of implementing
the groupings which access rights/users are assigned to
in the policy (roles in this paper) as a set 
of MVC components, and then statically verifying static groups whilst
dynamically verifying dynamic groups. The result
would allow static parts of the policy to be enforced statically, whilst
still allowing dynamic policies to be expressed and then enforced
dynamically.

 Furthermore, we will consider systems where a policy is defined as a
 combination of existing policies, extending the approach in order to
 allow programmers to combine validated RBAC implementations without
 re-doing all the static checks.

\bibliography{biblio}{}

\begin{thebibliography}{10}
\providecommand{\bibitemdeclare}[2]{}
\providecommand{\surnamestart}{}
\providecommand{\surnameend}{}
\providecommand{\urlprefix}{Available at }
\providecommand{\url}[1]{\texttt{#1}}
\providecommand{\href}[2]{\texttt{#2}}
\providecommand{\urlalt}[2]{\href{#1}{#2}}
\providecommand{\doi}[1]{doi:\urlalt{http://dx.doi.org/#1}{#1}}
\providecommand{\bibinfo}[2]{#2}

\bibitemdeclare{article}{basin:06:mds}
\bibitem{basin:06:mds}
\bibinfo{author}{David \surnamestart Basin\surnameend},
  \bibinfo{author}{J\"{u}rgen \surnamestart Doser\surnameend} \&
  \bibinfo{author}{Torsten \surnamestart Lodderstedt\surnameend}
  (\bibinfo{year}{2006}): \emph{\bibinfo{title}{Model Driven Security: From UML
  Models to Access Control Infrastructures}}.
\newblock {\sl \bibinfo{journal}{ACM Trans. Softw. Eng. Methodol.}}
  \bibinfo{volume}{15}(\bibinfo{number}{1}), pp. \bibinfo{pages}{39--91},
  \doi{10.1145/1125808.1125810}.

\bibitemdeclare{inproceedings}{BertolissiC:ppdp}
\bibitem{BertolissiC:ppdp}
\bibinfo{author}{Clara \surnamestart Bertolissi\surnameend} \&
  \bibinfo{author}{Maribel \surnamestart Fern\'{a}ndez\surnameend}
  (\bibinfo{year}{2008}): \emph{\bibinfo{title}{A Rewriting Framework for the
  Composition of Access Control Policies}}.
\newblock In: {\sl \bibinfo{booktitle}{Proceedings of the 10th International
  ACM SIGPLAN Conference on Principles and Practice of Declarative
  Programming}}, \bibinfo{series}{PPDP '08}, \bibinfo{publisher}{ACM},
  \bibinfo{address}{New York, NY, USA}, pp. \bibinfo{pages}{217--225},
  \doi{10.1145/1389449.1389476}.

\bibitemdeclare{article}{bodden:12:clara}
\bibitem{bodden:12:clara}
\bibinfo{author}{Eric \surnamestart Bodden\surnameend},
  \bibinfo{author}{Patrick \surnamestart Lam\surnameend} \&
  \bibinfo{author}{Laurie \surnamestart Hendren\surnameend}
  (\bibinfo{year}{2012}): \emph{\bibinfo{title}{Partially Evaluating
  Finite-State Runtime Monitors Ahead of Time}}.
\newblock {\sl \bibinfo{journal}{ACM Trans. Program. Lang. Syst.}}
  \bibinfo{volume}{34}(\bibinfo{number}{2}), pp. \bibinfo{pages}{7:1--7:52},
  \doi{10.1145/2220365.2220366}.

\bibitemdeclare{incollection}{BonattiS03}
\bibitem{BonattiS03}
\bibinfo{author}{Piero~A. \surnamestart Bonatti\surnameend} \&
  \bibinfo{author}{Pierangela \surnamestart Samarati\surnameend}
  (\bibinfo{year}{2004}): \emph{\bibinfo{title}{Logics for Authorizations and
  Security}}.
\newblock In \bibinfo{editor}{Jan \surnamestart Chomicki\surnameend},
  \bibinfo{editor}{Ron \surnamestart van~der Meyden\surnameend} \&
  \bibinfo{editor}{Gunter \surnamestart Saake\surnameend}, editors: {\sl
  \bibinfo{booktitle}{Logics for Emerging Applications of Databases}},
  \bibinfo{publisher}{Springer Berlin Heidelberg}, pp.
  \bibinfo{pages}{277--323}, \doi{10.1007/978-3-642-18690-5_8}.

\bibitemdeclare{article}{bovet:08:antlrworks}
\bibitem{bovet:08:antlrworks}
\bibinfo{author}{Jean \surnamestart Bovet\surnameend} \&
  \bibinfo{author}{Terence \surnamestart Parr\surnameend}
  (\bibinfo{year}{2008}): \emph{\bibinfo{title}{ANTLRWorks: An ANTLR Grammar
  Development Environment}}.
\newblock {\sl \bibinfo{journal}{Softw. Pract. Exper.}}
  \bibinfo{volume}{38}(\bibinfo{number}{12}), pp. \bibinfo{pages}{1305--1332},
  \doi{10.1002/spe.v38:12}.

\bibitemdeclare{book}{buschmann:96:posa}
\bibitem{buschmann:96:posa}
\bibinfo{author}{Frank \surnamestart Buschmann\surnameend},
  \bibinfo{author}{Regine \surnamestart Meunier\surnameend},
  \bibinfo{author}{Hans \surnamestart Rohnert\surnameend},
  \bibinfo{author}{Peter \surnamestart Sommerlad\surnameend} \&
  \bibinfo{author}{Michael \surnamestart Stal\surnameend}
  (\bibinfo{year}{1996}): \emph{\bibinfo{title}{Pattern-oriented Software
  Architecture: A System of Patterns}}.
\newblock \bibinfo{publisher}{John Wiley \& Sons, Inc.}, \bibinfo{address}{New
  York, NY, USA}.

\bibitemdeclare{inproceedings}{fer:92:rbac}
\bibitem{fer:92:rbac}
\bibinfo{author}{David \surnamestart Ferraiolo\surnameend} \&
  \bibinfo{author}{Richard \surnamestart Kuhn\surnameend}
  (\bibinfo{year}{1992}): \emph{\bibinfo{title}{Role-Based Access Control}}.
\newblock In: {\sl \bibinfo{booktitle}{In 15th NIST-NCSC National Computer
  Security Conference}}, pp. \bibinfo{pages}{554--563}.

\bibitemdeclare{article}{fer:01:rbacstandard}
\bibitem{fer:01:rbacstandard}
\bibinfo{author}{David~F. \surnamestart Ferraiolo\surnameend},
  \bibinfo{author}{Ravi \surnamestart Sandhu\surnameend},
  \bibinfo{author}{Serban \surnamestart Gavrila\surnameend},
  \bibinfo{author}{D.~Richard \surnamestart Kuhn\surnameend} \&
  \bibinfo{author}{Ramaswamy \surnamestart Chandramouli\surnameend}
  (\bibinfo{year}{2001}): \emph{\bibinfo{title}{Proposed NIST Standard for
  Role-based Access Control}}.
\newblock {\sl \bibinfo{journal}{ACM Trans. Inf. Syst. Secur.}}
  \bibinfo{volume}{4}(\bibinfo{number}{3}), pp. \bibinfo{pages}{224--274},
  \doi{10.1145/501978.501980}.

\bibitemdeclare{misc}{eclipse}
\bibitem{eclipse}
\bibinfo{author}{The~Eclipse \surnamestart Foundation\surnameend}:
  \emph{\bibinfo{title}{Eclipse}}.
\newblock \urlprefix\url{http://www.eclipse.org}.

\bibitemdeclare{book}{gosling:05:jls}
\bibitem{gosling:05:jls}
\bibinfo{author}{James \surnamestart Gosling\surnameend}, \bibinfo{author}{Bill
  \surnamestart Joy\surnameend}, \bibinfo{author}{Guy \surnamestart
  Steele\surnameend} \& \bibinfo{author}{Gilad \surnamestart Bracha\surnameend}
  (\bibinfo{year}{2005}): \emph{\bibinfo{title}{Java(TM) Language
  Specification, The (3rd Edition) (Java (Addison-Wesley))}}.
\newblock \bibinfo{publisher}{Addison-Wesley Professional}.

\bibitemdeclare{book}{gupta:jee}
\bibitem{gupta:jee}
\bibinfo{author}{Arun \surnamestart Gupta\surnameend} (\bibinfo{year}{2013}):
  \emph{\bibinfo{title}{Java EE 7 Essentials}}.
\newblock \bibinfo{publisher}{O'Reilly Media}.

\bibitemdeclare{article}{hamlen:06:ccem}
\bibitem{hamlen:06:ccem}
\bibinfo{author}{Kevin~W. \surnamestart Hamlen\surnameend},
  \bibinfo{author}{Greg \surnamestart Morrisett\surnameend} \&
  \bibinfo{author}{Fred~B. \surnamestart Schneider\surnameend}
  (\bibinfo{year}{2006}): \emph{\bibinfo{title}{Computability Classes for
  Enforcement Mechanisms}}.
\newblock {\sl \bibinfo{journal}{ACM Trans. Program. Lang. Syst.}}
  \bibinfo{volume}{28}(\bibinfo{number}{1}), pp. \bibinfo{pages}{175--205},
  \doi{10.1145/1111596.1111601}.

\bibitemdeclare{article}{krasner:88:mvc}
\bibitem{krasner:88:mvc}
\bibinfo{author}{Glenn~E. \surnamestart Krasner\surnameend} \&
  \bibinfo{author}{Stephen~T. \surnamestart Pope\surnameend}
  (\bibinfo{year}{1988}): \emph{\bibinfo{title}{A Cookbook for Using the
  Model-view Controller User Interface Paradigm in Smalltalk-80}}.
\newblock {\sl \bibinfo{journal}{J. Object Oriented Program.}}
  \bibinfo{volume}{1}(\bibinfo{number}{3}), pp. \bibinfo{pages}{26--49}.
\newblock \urlprefix\url{http://dl.acm.org/citation.cfm?id=50757.50759}.

\bibitemdeclare{inproceedings}{priebe:04:pattern}
\bibitem{priebe:04:pattern}
\bibinfo{author}{Torsten \surnamestart Priebe\surnameend},
  \bibinfo{author}{Eduardo~B. \surnamestart Fernandez\surnameend},
  \bibinfo{author}{Jens~I. \surnamestart Mehlau\surnameend} \&
  \bibinfo{author}{G{\"u}nther \surnamestart Pernul\surnameend}
  (\bibinfo{year}{2004}): \emph{\bibinfo{title}{A pattern system for access
  control}}.
\newblock In: {\sl \bibinfo{booktitle}{Research Directions In Data and
  Applications Security XVIII}}, \bibinfo{publisher}{Kluwer}, pp.
  \bibinfo{pages}{25--28}, \doi{10.1007/1-4020-8126-6_16}.

\bibitemdeclare{phdthesis}{SantanaA:phd}
\bibitem{SantanaA:phd}
\bibinfo{author}{A.~\surnamestart {Santana de Oliveira}\surnameend}
  (\bibinfo{year}{2008}): \emph{\bibinfo{title}{{\em R\'e\'ecriture et
  Modularit\'e pour les Politiques de S\'ecurit\'e}}}.
\newblock Ph.D. thesis, \bibinfo{school}{Universit\'e Henri Poincare},
  \bibinfo{address}{Nancy, France}.

\bibitemdeclare{article}{SohrK:Isabelle}
\bibitem{SohrK:Isabelle}
\bibinfo{author}{Karsten \surnamestart Sohr\surnameend},
  \bibinfo{author}{Michael \surnamestart Drouineaud\surnameend},
  \bibinfo{author}{Gail-Joon \surnamestart Ahn\surnameend} \&
  \bibinfo{author}{Martin \surnamestart Gogolla\surnameend}
  (\bibinfo{year}{2008}): \emph{\bibinfo{title}{Analyzing and Managing
  Role-Based Access Control Policies}}.
\newblock {\sl \bibinfo{journal}{IEEE Transactions on Knowledge and Data
  Engineering}} \bibinfo{volume}{20}(\bibinfo{number}{7}), pp.
  \bibinfo{pages}{924--939}, \doi{10.1109/TKDE.2008.28}.

\bibitemdeclare{book}{steel:06:csp}
\bibitem{steel:06:csp}
\bibinfo{author}{Christopher \surnamestart Steel\surnameend},
  \bibinfo{author}{Ramesh \surnamestart Nagappan\surnameend} \&
  \bibinfo{author}{Ray \surnamestart Lai\surnameend} (\bibinfo{year}{2006}):
  \emph{\bibinfo{title}{{Core security patterns: Best practices and strategies
  for J2EE, Web services, and identity management}}}.
\newblock \bibinfo{series}{Prentice Hall Core Series},
  \bibinfo{publisher}{Prentice-Hall}.
\newblock \urlprefix\url{http://www.coresecuritypatterns.com/}.

\bibitemdeclare{inproceedings}{zarnett:10:annot}
\bibitem{zarnett:10:annot}
\bibinfo{author}{Jeff \surnamestart Zarnett\surnameend},
  \bibinfo{author}{Mahesh \surnamestart Tripunitara\surnameend} \&
  \bibinfo{author}{Patrick \surnamestart Lam\surnameend}
  (\bibinfo{year}{2010}): \emph{\bibinfo{title}{Role-based Access Control
  (RBAC) in Java via Proxy Objects Using Annotations}}.
\newblock In: {\sl \bibinfo{booktitle}{Proceedings of the 15th ACM Symposium on
  Access Control Models and Technologies}}, \bibinfo{series}{SACMAT '10},
  \bibinfo{publisher}{ACM}, \bibinfo{address}{New York, NY, USA}, pp.
  \bibinfo{pages}{79--88}, \doi{10.1145/1809842.1809858}.

\end{thebibliography}
\bibliographystyle{eptcs}

\end{document}